# Micromagnetic Simulations for Spin Transfer Torque in Magnetic Multilayers

Chun-Yeol You[1]

[1]Department of Physics, Inha University, Incheon 402-751, Korea

We investigate the spin transfer torque (STT) in the magnetic multilayer structures with micromagnetic simulations. We implement the STT contribution for the magnetic multilayer structures in addition to the Landau-Lifshitz–Gilbert (LLG) micromagnetic simulators. Not only the Sloncewski STT term, the zero, first, and second order field-like terms are also considered, and the effects of the Oersted field by the current are addressed. We determine the switching current densities of the free layer with the exchange biased synthetic ferrimagnetic reference layers for various cases.

## 1. Introductions

In the theoretical study of spin transfer torque (STT), there are three categories. The main concern of the first one is finding the physical origin of the STT in a given system. The goals of the simple free electron models [1,2], first principle calculations [3], and Keldysh non-equilibrium Green's function methods [4,5,6,7] are finding the physical origin of the STT term in the given systems. Such study reveals the existence of the STT term and what kinds of material parameters govern the magnitude of the STT. The second category is the study of spin dynamics with simple macro spin models. By analytically or numerically solving the Landau-Lifshitz–Gilbert (LLG) equation with the additional STT term, switching current density can be determined in the given systems [8,9], and it is helpful to get rough idea about the spin dynamics under the STT. However, in order to investigate the details of the spin dynamics, the macro spin model is too simple and the micromagnetic approaches are essential. By the virtue of the open source micromagnetic simulation, OOMMF (Object Oriented MicroMagnetic Frame) [10] and the implementation of the STT in the nanowire geometry [11], many researches about the STT in the domain wall motion in the nanowire have been reported [12]. However, so far, there is no open source micromagnetic simulator with STT term for the magnetic multilayer structures including exchange biased synthetic ferrimagnetic reference layers. In this study, we implement the add-on extension module for the STT in the magnetic multilayer structures. The developed extension module is based on the OOMMF, the OOMMF users can easily handle the STT effect in the nano-pillar geometry with magnetic multilayers, such as typical STT-MRAM (magnetoresistive random access memory) structures [13,14,15,16]. First, we will explain the details of the implementation of STT module and usages [17], and show the micromagnetic simulation results with various simulation parameters.

## 2. Implementation of STT term in OOMMF

We add the in-plane and out-of-plane STT terms in LLG equations to implement the STT module in the OOMMF.

$$\frac{d\vec{m}_s}{dt} = -\gamma \vec{m}_s \times \vec{H}_{eff} + \alpha \vec{m}_s \times \frac{d\vec{m}_s}{dt} - \gamma a_J \vec{m}_s \times (\vec{m}_s \times \vec{m}_p) - \gamma b_J (\vec{m}_s \times \vec{m}_p) \quad (1)$$

Here, $a_J = a_1 J$, $a_1 = \eta_p \frac{\hbar}{2e\mu_0 M_s d_s}$, $b_J = (b_0 + b_1 J + b_2 J^2)$ and $J$ is current density (opposite to the electron flow), and $\vec{m}_{s,p}$ are unit vectors of the magnetization of switching and polarizer layers, respectively. We consider the zero-th, first, and second order field-like terms in order to handle most general cases [14,15,16]. $\eta_p$ is spin polarization of the polarizer layer, or spin torque efficiency. We set $\eta_p$ is 0.7 for our study, and $d_s$ is the thickness of the switching layer. In the present implementation, the $a_1$ is automatically determined with the given parameters. According to the theoretical works [18], $\eta_p$ has angular variation, however, we ignore the angular dependence of $\eta_p$ in our implementation. It must be noted that we assume the STT is interface term, so that we replace $d_s$ with the unit cell thickness of z-direction in above relation. Therefore, the STT acts only on the first unit cells of the switching layer, and the next unit cells coupled with first unit cells only by the atomic exchange coupling. In this way, the thicker switching layer can be handled automatically. We will discuss the switching layer thickness dependence of the switching current density later.

The direction of positive current is defined from the bottom to top of the nano-pillar as shown in the Fig. 1. It must emphasized that the positive (negative) current prefer anti-parallel (parallel) states, because the electron flows are opposite to the current direction.

The Oersted field generated by the current in the nano-pillar structure is numerically calculated by the separate procedures, and the calculated Oersted field is read as an external field. The magnitude of the Oersted field is determined by the current density and nano-pillar geometry. It must be mentioned that we assume the uniform current density, even though the real current density is non-uniform during the switching processes due to the relatively large tunneling magneto-resistance (TMR) in MgO based junctions. With built-in exchange bias, and interlayer exchange coupling energy, the typical STT-MRAM structures, AFM (anti-ferromagnetic), synthetic ferrimagnet ($F_3$/NM/$F_2$), Insulator (I), and free ($F_1$) layers can be successfully modeled and examined in this study. Where, $F_{1,2,3}$ are ferromagnetic layers and NM is non-magnetic layer, respectively. More details of the usages can be found [17].

### 3. Results and Discussions

We consider typical STT-MRAM structures as shown in the Fig. 1. The saturation magnetization $M_s$ and thicknesses of $F_{1,\ 2,\ 3}$ layers are $1.3 \times 10^6$ A/m and 2 nm, respectively, except when we investigate the $F_1$ layer thickness dependence. The thicknesses of NM and I layers are 1 nm. The cross-section of the nano-pillar is ellipse of $60 \times 40$ nm$^2$, with the cell size of $1 \times 1 \times 1$ nm$^3$. The exchange bias field of $4 \times 10^5$ A/m is assigned to +$x$-direction for the $F_3$ layer. The strong antiferromagnetic interlayer exchange coupling energy $-1 \times 10^{-3}$ J/m$^2$ is applied between $F_2$ and $F_3$ layers in order to keep antiferromagnetic coupling between them. No crystalline anisotropy energy is considered in this study for the simplicity. The exchange stiffness constant *A* set as $2.0 \times 10^{-11}$ J/m, and the Gilbert damping constants $\alpha$ are varied from 0.005 to 0.05.

Fig. 2 shows hysteresis loop of the $F_1$ layer. Due to the stray field from the synthetic ferrimagnet ($F_3$/NM/$F_2$) structure, the loop is shifted by -17 kA/m from the center. The stray field causes the different switching current densities for P (parallel) to AP (anti-parallel)-state and AP to P-state switching. It must be noted that the micromagnetic simulation reveals that the stray field from the synthetic ferrimagnet at the $F_1$ layer position is varied from -1.2 to -31 kA/m (from center to edge). Therefore, the -17 kA/m shift is reasonable.

The switching current density $J_c$ is known as

$$J_c \sim \frac{\alpha}{a_1} \left[ H_{eff} + \frac{1}{2}(N_y + N_z - 2N_x) M_s \right], \qquad (2)$$

for the macro-spin model [8, 9]. Where $N_{x,y,z}$ are demagnetization factor of the $F_1$ layer, $N_z = 1$, $N_x = N_y = 0$ for the thin film, and $H_{eff} = H_{ext} + b_J + H_{stray} + H_{Oe}$ is an effective field including the external field, $H_{ext}$, field like term, $b_J$, stray field, $H_{stray}$, and Oersted field, $H_{Oe}$.

In order to determine the switching current density, we perform the micromagnetic simulation with the current density for 10 ns, and wait 2 ns more without current. After 12 ns simulation time, we check the magnetization configurations, and decide the switching status. With this procedure, first, we varied $F_1$ thickness, $d_s$, from 1 ~ 4 nm, and find $J_c$ for each thickness. The results are depicted in Fig. 3 for P- to AP-state switching. The symbols are micromagnetic simulation results and the solid line is a linear fit. As shown in Fig. 3, the dependence of the switching current density on $d_s$ is well fitted with a linear line, because the inverse proportionality of $a_1$ to $d_s$. Since we assume the STT acts only the interface, the first unit cell of switching layer, the linearity is what we expect. We find the linearity of broken when $d_s > 5$ nm in our simulations. In thicker $d_s$, the bottom of $F_1$ layer is switched where the spin torque is exerted, but the upper part of $F_1$ layer is not switched together due to the finite exchange length. Therefore, a twisted domain wall is formed between the bottom and top layer. Because the exchange length of switching layer is order of a few nm, it is reasonable results. However, if the domain wall is formed in real experiments, the additional spin torque is created due to the non-collinear alignment of the magnetization inside of the switching layer. And the additional spin torque will move the domain wall. However, such kind spin torque is not implemented in our simulations. Therefore, we have to limit our simulation to thin switching layer, and the limited thickness is determined by exchange length.

Second, we show the Gilbert damping parameter $\alpha$ dependence of the $J_c$ in Fig. 4. Equation (2) implies the switching current density is proportional to $\alpha$. The linear dependence is clearly shown in Fig. 4 for P- to AP-state and AP- to P-state, respectively. If we calculate the slope of the relation $J_c \sim D\alpha$, $D = \frac{1}{a_1}\left(\pm H_{stray} \pm H_{Oe} + \frac{1}{2}M_s\right)$ from the given parameters, we obtain $D_{P-AP} = 9.0$ and $D_{AP-P} = 9.5 \times 10^{12}$ A/m$^2$ without consideration of the Oersted field. The slopes from the micromagnetic simulations are $D_{P-AP} = 11.9$ and $|D_{AP-P}| = 12.4 \times 10^{12}$ A/m$^2$. Even though there are small discrepancies in the absolute magnitudes between the analytic and micromagnetic results, the difference between P- to AP-state and AP- to P-state switching current density, $D_{AP-P} - D_{P-AP}$, are similar for both cases, and it implies the origin of the difference switching current density is the stray field [19]. The small discrepancies between analytic expression and micromagnetic simulation indicate the limits of the macro-spin model. The macro-spin model ignores the spin wave excitation with the short wavelength (see Fig. 5 (b) and (c)). Since the short wavelength spin wave excitation is important in the switching procedure, the analytic expression cannot be correct. One more finding in our simulation results are the finite interception when $\alpha \to 0$. According to the analytic model, the $J_c$ goes zero with $\alpha$. However, the micromagnetic simulation reveals $J_c$ has the finite values when $\alpha \to 0$. The physical reason of the finite interception is not clear in this stage. Even though the values are not large (-7.32 and $3.74 \times 10^{10}$ A/m$^2$ for AP- to P-state and P- to AP-state, respectively), it may be important to reduce the $J_c$ for small $\alpha$ materials.

Fig. 5 (a) ~ (d) shows the snap shots of the $F_1$ layer magnetization configurations during switching processes. With the current density of $2.7 \times 10^{11}$ A/m$^2$ and $\alpha$ of 0.02, the magnetization of $F_1$ layer switches from P-state to AP-state at $t = 9.0$ ns as shown in Fig. 5 (e). According to Fig. 5 (e), the magnetization oscillates after 2 ns, but it takes 6~7 ns till the switching occurs. During that period, the magnetization oscillates as shown in Fig. 5 (a). The spins oscillate and the directions of the spin are opposite at both edges. C-type domain structure is formed, and it implies the macro-spin model cannot describe the Fig. 5 (a) state. At $t = 7.7$ ns, the spin dynamics is getting more complex as shown in Fig. 5 (b). The domain structure is S-type, and the instability of the domain structure increases with the spin polarized current. At $t = 8.8$ ns, Fig 5 (c), most spins point short axis of the ellipse and finally the complete switching occurs after 9.0 ns, Fig. 5 (d). With different $\alpha$ gives similar spin dynamics behaviors.

We include or exclude the numerically pre-calculate the Oersted field which is generated by the current in the simulations. The effect of the Oersted field is not serious in the determination of the switching current density. However, the Oersted field changes the details of the spin dynamics. Figure 6 shows the magnetization switching from AP- to P-state with the current density of $-1.84 \times 10^{11}$ A/m$^2$ with and without the Oersted field for $\alpha = 0.01$ cases. As shown in Fig. 6, the switching occurs without the Oersted field, while the switching does not occurs with the Oersted field. However, the difference of the switching current density is not noticeable in many cases.

It must be mentioned that we do not show the effect of the field like term in this work. Since the role of the field-like term is tedious, but the proper field-like term contribution is not clear yet [13,14,15,16]. However, it is already implemented in our extension module, there is no limitation of the study of the field-like term contribution.

## 4. Conclusions

We implement the STT extension module for the OOMMF. With the STT extension module, we simulate the spin dynamics for typical STT-MRAM structures, consist of the exchange biased synthetic ferrimagnetic layer and free layer with the insulator layer. We find that the switching current densities for P- to AP-state and AP- to P-state are different even though we set the $a_1$ are identical for both case. The small difference of the switching current densities is ascribed stray field from the exchange biased synthetic ferrimagnetic layer.

### Acknowledgements

This work was supported by (2010-0022040) programs through a NRF grant funded by MEST.

**Figure Captions**

Fig. 1 The typical STT-MRAM structure. $F_{1,2,3}$ are the ferromagnetic and NM, I, AFM are the non-magnetic, insulator, antiferromagnetic layers, respectively. The positive current density is defined from the bottom to top, and $x$-axis is long axis of the ellipse.

Fig. 2 Magnetization hysteresis loop of the $F_1$ layer. The hysteresis loop is shifted by the stray field from the synthetic ferrimagnet structure ($F_3$/NM/$F_2$). Within the swept field range, the magnetization of $F_2$ and $F_3$ layers are almost fixed.

Fig. 3 The switching current densities $J_c$ as a function of the switching layer thickness for P- to AP-state switching. The symbols are micromagnetic simulation results and solid line is a linear fit.

Fig. 4 The switching current densities $J_c$ as a function of the Gilbert damping parameter $\alpha$ for P- to AP-state (blue symbols) and AP- to P-state (red symbols). The solid lines are linear fits.

Fig. 5 The snap shots of the magnetization configuration of the $F_1$ layer with $J = 2.7 \times 10^{11}$ A/m$^2$ at (a) $t = 5.0$, (b) 7.7, (c) 8.8 and (d) 9.0 ns, respectively. (e) The magnetization dynamics as a function of the time for $\alpha = 0.02$.

Fig. 6 The magnetization dynamics from AP- to P- state as a function of the time for $\alpha = 0.01$ with and without the Oersted field generated by the spin polarized current of $-1.84 \times 10^{11}$ A/m$^2$.

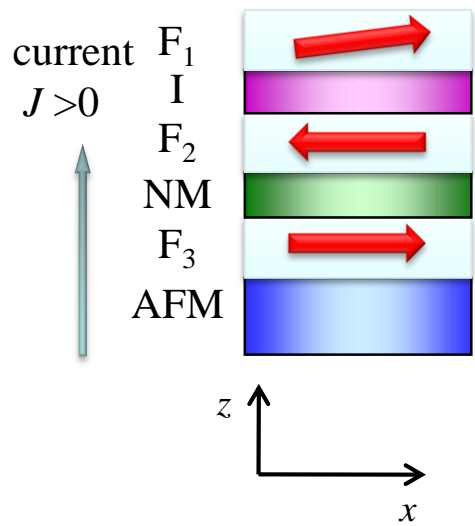

Fig. 1

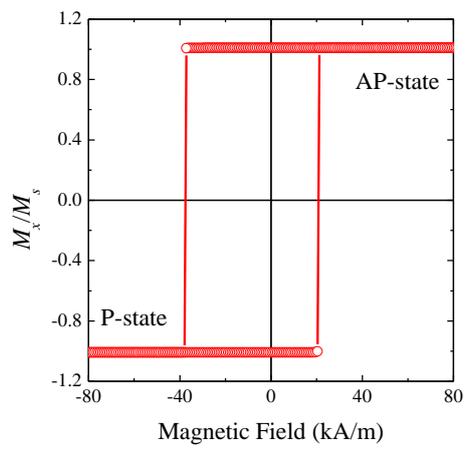

Fig. 2

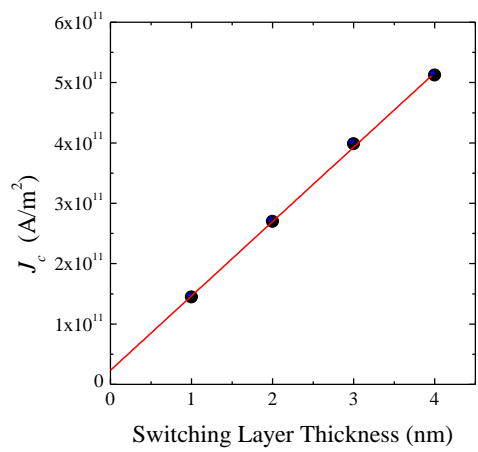

Fig. 3

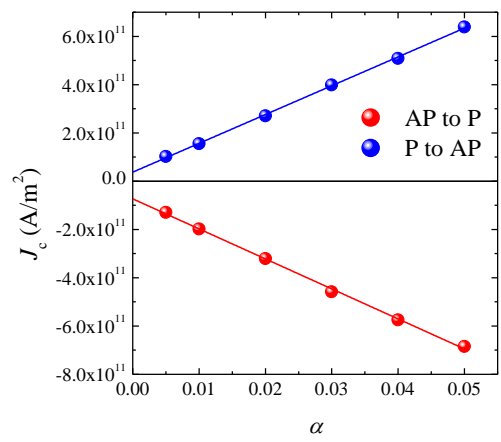

Fig. 4

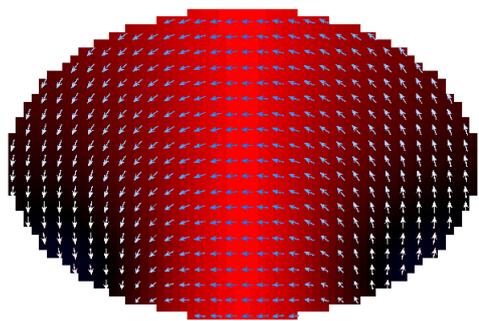 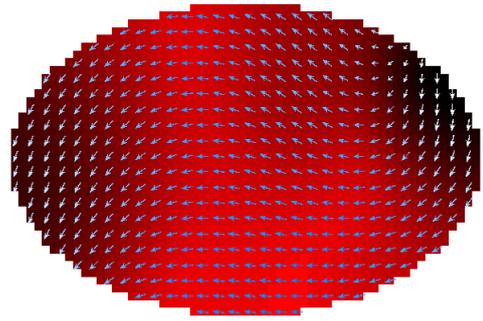

(a) $t = 5.0$ ns  (b) $t = 7.7$ ns

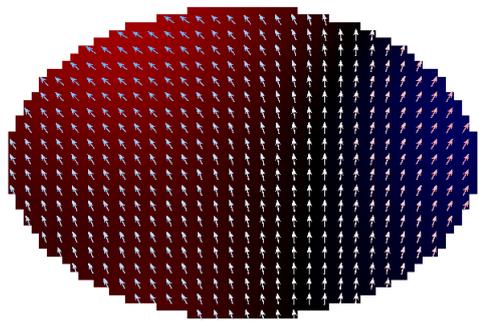 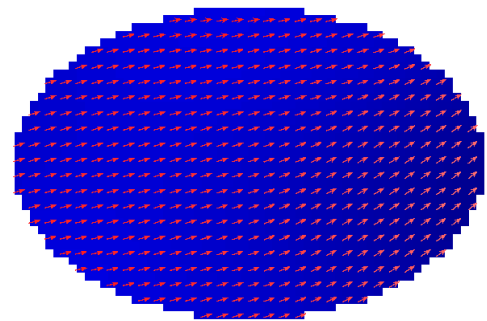

(c) $t = 8.8$ ns  (d) $t = 9.0$ ns

Fig. 5 (a)~ (d)

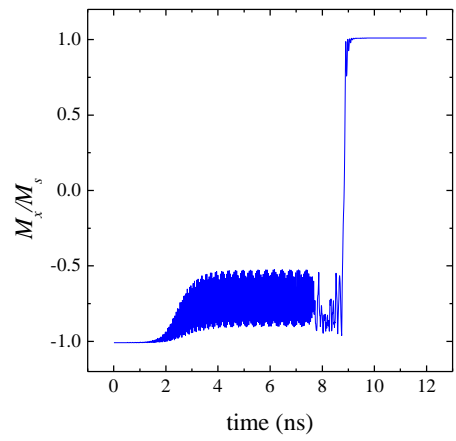

Fig. 5 (e)

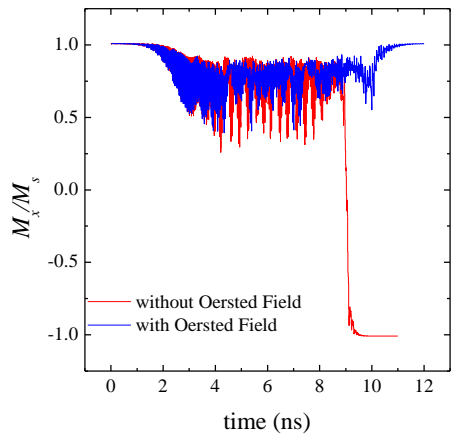

Fig. 6